\documentclass[pss]{wiley2sp}
\usepackage{amsmath}
\usepackage{bm}
\usepackage{w-greek}
\usepackage{color}

\begin{document}

\title{Magnetic properties of the spin-1/2 Ising-Heisenberg diamond chain with the four-spin interaction}

\titlerunning{Diamond chain with the four-spin interaction}

\author{%
Lucia G\'alisov\'a}
\authorrunning{L. G\'alisov\'a}

\mail{e-mail
    \textsf{galisova.lucia@gmail.com}, Phone: + 421 55 602 2228}

\institute{%
    Department of Applied Mathematics and Informatics, Faculty of Mechanical Engineering, Technical University, Letn\'a 9, 042 00 Ko\v{s}ice, Slovak Republic}

\received{14 October 2007, revised XXXX, accepted XXXX}
\published{XXXX}

\keywords{Ising-Heisenberg diamond chain, four-spin interaction, decoration-iteration transformation, exact results}%

\abstract{%
\abstcol{%
A symmetric spin-1/2 Ising-Heisenberg diamond chain with the Ising four-spin interaction is exactly solved by means of the generalized decoration-iteration mapping transformation. The ground state, the magnetization process and thermodynamics are particularly examined for the case of antiferromagnetic pair interactions (Ising and isotropic Heisenberg ones). It is shown that an interplay between pair interactions, the four-spin interaction and}{%
the external magnetic field gives rise to several quantum ground states with entangled spin states in addition to some semi-classically ordered ones. Besides, the temperature dependence of the magnetic susceptibility multiplied by the temperature is studied and the interesting triple-peak specific heat curve is also detected when considering the zero-field region rather close to the triple point, where three different ground states coexist.
}}

\maketitle

\section{Introduction}
Spin systems with multispin exchange interactions represent objects of scientific interest in the past few years. Research in this field leads to a deeper understanding of many interesting physical phenomena, such as the non-universal critical behaviour~\cite{Wu71,Kad71}, optical conductivity~\cite{Nun02}, Raman peaks~\cite{Sch01}, as well as, deviations from the Bloch $T^{3/2}$ law at low temperatures~\cite{Kob96,Mul97}. Moreover, the effect of magnetic field on the ground state of quantum systems with cyclic four-spin interaction has been recently particularly examined as well~\cite{Nak01,Hik08a,Hik08b}. Of course, the immense theoretical interest to the spin models with multispin interactions is not purposeless. Multispin interactions have been experimentally observed in real triangular magnetic systems composed of $^3$He atoms absorbed on graphite surfaces~\cite{Rog83}, the hydrogen bonded ferroelectrics PbHPO$_4$ and PbDPO$_4$~\cite{Chu88}, as well as, the squaric acid crystal (H$_2$C$_2$O$_4$)~\cite{Wan80,Wan89,Wan90} and some copolymers~\cite{Sil93}. Recently, it was shown that the cyclic four-spin interaction could also explain the neutron-scattering experiments concerning high-$T_c$ compounds such as La$_2$CuO$_4$~\cite{Col01}, La$_6$Ca$_8$Cu$_{24}$O$_{41}$~\cite{Bre99,Mat00}, and La$_4$Sr$_{10}$Cu$_{24}$O$_{41}$~\cite{Not07}.

On the theoretical side, in spite of the numerous numerical studies predicting rich phase diagrams~\cite{Sch03,Lau03,Lech06,Iva09},
\begin{figure*}[h]%
\vspace{-1.0cm}
  \sidecaption
    \includegraphics*[width=.5\textwidth,height=5.20cm]{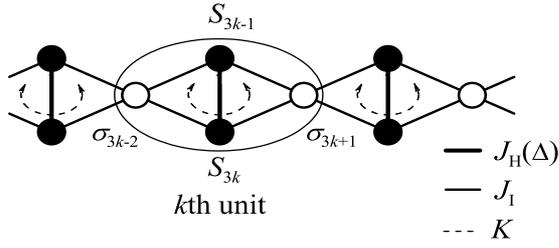}%
    \vspace{0.0cm}
    \caption{%
    A part of the spin-1/2 Ising-Heisenberg diamond chain with the four-spin interaction. The empty (full) circles denote lattice positions of the Ising (Heisenberg) spins. The ellipse demarcates spins belonging to the $k$th diamond unit.}
    \label{fig1}
\end{figure*}
a number of important questions concerning the spin ordering and the critical behaviour of quantum Heisenberg models realized by the four-spin interaction, remain unclear due to controversial results obtained by different treatments~\cite{Sen04,San07,Kot09,Isa10}. From this point of view, the exactly solvable models play an important role in understanding the multispin effects. Of course, the quantum spin models is very difficult to deal with exactly due to a rather cumbersome and sophisticated mathematics, which precludes an exact treatment of the most (even simple-minded) spin systems. Motivated by this fact, a special class of the hybrid {\it Ising-Heisenberg models}~\cite{Str02a,Str02b,Str03,Oha03,Can04,Str05,Str06,Can06,Can07,Str08,Oha09,Ant09,Oha10,Kiss,Str,Oha} has been recently proposed, which overcome the afore-mentioned mathematical difficulty by introducing the Ising spins at nodal lattice sites and the Heisenberg dimers on interstitial decorating sites of the considered lattice. These simplified classical-quantum spin models can be rigorously solved by two different exact analytical approaches; the classical transfer-matrix technique~\cite{Kra44,Bax82}, which is rather straightforward but useable only for one-dimensional spin systems, and/or more universal generalized decoration-iteration transformation~\cite{Fis59,Syo72,Roj09,Str10}, which can be used for both the one- and two-dimensional systems. It is also worth mentioning that in order to investigate many interesting physical phenomena, various extensions and generalizations of the hybrid Ising-Heisenberg models may be done without loss of their exact solubility. For instance, it is possible to extend the Ising-Heisenberg models by including the next-nearest-neighbour exchange interaction between the Ising spins~\cite{Val08}, the Dzyaloshinskii-Moriya anisotropy acting on the decorating Heisenberg spins~\cite{Str09b},
or to solve exactly the analogous Ising-Heisenberg models with the Heisenberg spins $S>1/2$~\cite{Str02b,Can06,Ayd05,Ayd06,Can08,Jas08,Can09,Str09a,Can10,Gal11,Roj11,Bel10} that bring a deeper insight into how the magnetic behaviour of the quantum spin systems depends on the magnitude $S$~\cite{Can06,Can09,Str09a} of the decorating Heisenberg spins and also allow to examine the effect of other interaction terms such as the axial zero-field splitting parameter~\cite{Str02b,Ayd05,Ayd06,Can08,Jas08,Can10,Gal11,Roj11} and/or the biquadratic XXZ interaction~\cite{Roj11,Bel10}.

In this work, we will investigate the symmetric spin-1/2 Ising-Heisenberg diamond chain, which extends the class of lattice-statistical models considered in Refs.~\cite{Can04,Can06} by including the Ising four-spin interaction. Beside the ground-state analysis, our main goal is to particularly examine the effect of the four-spin interaction on the magnetization scenario as well as the thermodynamics of the Ising-Heisenberg diamond chain. Note that the considered multispin interaction does not violate the standard commutation relations for cluster Hamiltonians of two different diamond-shaped units, and hence all the results for the investigated extended version of the Ising-Heisenberg diamond chain will be attained by combining the generalized decoration-iteration mapping transformation~\cite{Fis59,Syo72,Roj09,Str10} and the transfer-matrix technique~\cite{Kra44,Bax82}.

The organization of this paper is as follows. In Section~\ref{sec:model}, we will define the symmetric spin-1/2 Ising-Heisenberg diamond chain with the four-spin exchange interaction and then, the most important steps of an exact analytical treatment will be reviewed. Section~\ref{sec:results} deals with a discussion of the most interesting numerical results for the ground state, the magnetization process, the susceptibility and the specific heat. Finally, some concluding remarks are drawn in Section~\ref{sec:conclusions}.

\section{Model and its exact solution \label{sec:model}}
Let us consider an one-dimensional lattice of $N$ inter-connected diamonds (see Fig.~\ref{fig1}) defined by the Hamiltonian $\hat{{\cal H}} = \sum_{k=1}^{N}\hat{{\cal H}}_k$, where
\begin{eqnarray}
\label{eq:Hk}
\hat{{\cal H}}_k &=& J_{\rm H}\!\left[\Delta(\hat{S}_{3k-1}^{x}\hat{S}_{3k}^{x} + \hat{S}_{3k-1}^{y}\hat{S}_{3k}^{y}) + \hat{S}_{3k-1}^{z}\hat{S}_{3k}^{z}\right]\! \nonumber \\
&+& J_{\rm I}(\hat{S}_{3k-1}^{z}+\hat{S}_{3k}^{z})(\hat{\sigma}_{3k-2}^{z}+\hat{\sigma}_{3k+1}^{z}) {} \nonumber \\
&+& K \hat{S}_{3k-1}^{z}\hat{S}_{3k}^{z}\hat{\sigma}_{3k-2}^{z}\hat{\sigma}_{3k+1}^{z}{}
- H_{\rm H}(\hat{S}_{3k-1}^{z}+\hat{S}_{3k}^{z})
\nonumber\\
&-& H_{\rm I}(\hat{\sigma}_{3k-2}^{z}+\hat{\sigma}_{3k+1}^{z})/2.
\end{eqnarray}
Here, $\hat{S}_{k}^{\gamma}$ ($\gamma = x,y,z$) and $\hat{\sigma}_{k}^{z}$ denote spatial components of the spin-$1/2$ operators, the parameter $J_{\rm H}(\Delta)$ labels the anisotropic $XXZ$ interaction between the nearest-neighbouring Heisenberg spins, in which the parameter $\Delta$ allows to control the interaction $J_{\rm H}$ between the easy-axis ($\Delta<1$) and easy-plane ($\Delta>1$) type, as well as, to obtain the Ising model as a special limiting case when assuming $\Delta=0$. The parameter $J_{\rm I}$ denotes the Ising interaction between the Heisenberg spins and their nearest Ising neighbours, and the parameter $K$ describes the Ising four-spin interaction between both Heisenberg spins and two Ising spins of the diamond-shaped unit. Finally, the last two terms determine the magnetostatic Zeeman's energy of the Ising and Heisenberg spins placed in an external magnetic field $H_{\rm I}$ and $H_{\rm H}$ oriented along the $z$-axis, respectively.

Exact solution for the considered Ising-Heisenberg diamond chain can be achieved by following the same procedure as in our recent work~\cite{Can06}, where we have solved exactly the simplified version of this model (the model given by the Hamiltonian~(\ref{eq:Hk}), but without the term describing the four-spin interaction). Indeed, by taking into account a validity of the commutation relation between the cluster Hamiltonians of different diamond units, the partition function ${\cal Z}$ of the investigated Ising-Heisenberg diamond chain can be partially factorized:
\begin{eqnarray}
\label{eq:Z_IH}
{\cal Z} = \sum_{\{\sigma_k\}}{\rm Tr}_{\{S_k\}}\,{\rm e}^{-\beta\hat{{\cal H}}}= \sum_{\{\sigma_k\}}\prod_{k=1}^{N}{\rm Tr}_k\,{\rm e}^{-\beta\hat{{\cal H}}_k}.
\end{eqnarray}
Here, $\beta = 1/(k_{\rm B}T)$ ($k_{\rm B}$ is Boltzmann's constant, $T$ is the absolute temperature) and $N$ is the total number of the Ising spins. The symbols $\sum_{\{\sigma_k\}}$ and ${\rm Tr}_{\{S_k\}}$ denote a summation over spin degrees of freedom of all Ising and Heisenberg spins, respectively, while ${\rm Tr}_k$ means the partial trace over spin degrees of freedom of two Heisenberg spins from $k$th diamond unit.
It is obvious from the structure of the partition function (\ref{eq:Z_IH}) that
it is necessary to compute the last partial trace ${\rm Tr}_k{\rm e}^{-\beta\hat{{\cal H}}_k}$ in order to proceed further with a calculation. For this purpose, it is useful to pass to the matrix representation of the Hamiltonian~(\ref{eq:Hk}) in the orthogonal basis constructed by the eigenstates of the operator $\hat{S}_{3k-1}^{z}\hat{S}_{3k}^{z}$:
$\{|+, +\rangle_{3k-1, 3k}$, $|+, -\rangle_{3k-1, 3k}$, $|-, +\rangle_{3k-1, 3k}$, $|-, -\rangle_{3k-1, 3k}\}$, where the relevant ket vectors $|+(-), +(-)\rangle_{3k-1, 3k} = |+(-)\rangle_{3k-1}|+(-)\rangle_{3k}$ correspond to the appropriate combinations of the spin states $S_{3k-1}^{z} = \pm 1/2$ and $S_{3k}^{z} = \pm 1/2$ of two Heisenberg spins at $(3k-1)$th and $3k$th decorating sites, respectively. The diagonalization of the Hamiltonian~(\ref{eq:Hk}) transcribed in the matrix representation yields the four energy eigenvalues:
\begin{eqnarray}
\label{eq:E_k1}
{\cal E}_{1}(\sigma_{3k-2}^{z},\sigma_{3k+1}^{z}) &=& J_{\rm H}/4 + K\sigma_{3k-2}^{z}\sigma_{3k+1}^{z}/4 - H_{\rm H} \nonumber\\
&+&\left(J_{\rm I} - H_{\rm I}/2\right)\!(\sigma_{3k-2}^{z}\!+\sigma_{3k+1}^{z})  ,\\
\label{eq:E_k2}
{\cal E}_{2}(\sigma_{3k-2}^{z},\sigma_{3k+1}^{z}) &=& J_{\rm H}/4 + K\sigma_{3k-2}^{z}\sigma_{3k+1}^{z}/4 + H_{\rm H} \nonumber\\
&-& \left(J_{\rm I} + H_{\rm I}/2\right)\!(\sigma_{3k-2}^{z}\!+\sigma_{3k+1}^{z})  ,\\
\label{eq:E_k34}
{\cal E}_{3,4}(\sigma_{3k-2}^{z},\sigma_{3k+1}^{z}) &=& -J_{\rm H}/4 - K\sigma_{3k-2}^{z}\sigma_{3k+1}^{z}/4 \nonumber\\
&\pm& J_{\rm H}\Delta/2 - H_{\rm I}(_{3k-2}^{z}\!+\sigma_{3k+1}^{z})\!/2.
\end{eqnarray}
The eigenenergies (\ref{eq:E_k1})--(\ref{eq:E_k34}) can be straightforwardly used to obtain the last trace emerging in the expression of the partition function (\ref{eq:Z_IH}).
The resulting expression immediately implies the possibility of performing the generalized decoration-iteration mapping transformation given by Eq.~(3) in Ref.~\cite{Can06} with the modified function $G[\beta J_{\rm I}(\sigma_{3k-2}^{z}+\sigma_{3k+1}^{z}),\beta K\sigma_{3k-2}^{z}\sigma_{3k+1}^{z}]$:
\begin{eqnarray}
\label{eq:G}
G(x,y) &=& 2{\rm e}^{-\beta J_{\rm H}/4 - y/4}\cosh(x - \beta H_{\rm H}) \nonumber\\
&&{}
+ 2{\rm e}^{\,\beta J_{\rm H}/4 + y/4}\cosh(\beta J_{\rm H}\Delta/2).
\end{eqnarray}
Finally, a direct substitution of the mapping transformation given by Eq.~(3) in Ref.~\cite{Can06} into the expression~(\ref{eq:Z_IH}) yields to the simple relation between the partition function ${\cal Z}$ of the investigated mixed-spin Ising-Heisenberg diamond chain and the partition function ${\cal Z}_{\rm IC}$ of the uniform spin-$1/2$ Ising linear chain with the nearest-neighbour coupling $R$ and the effective magnetic field $H_{\rm IC}$
\begin{eqnarray}
\label{eq:Z}
{\cal Z}(\beta, J_{\rm I}, J_{\rm H}, K, \Delta, H_{\rm I}, H_{\rm H}) = A^{N}{\cal Z}_{\rm IC}(\beta, R, H_{\rm IC}).
\end{eqnarray}
The mapping parameters $A$, $R$ and $H_{\rm IC}$ emerging in Eq.~(\ref{eq:Z}) can be obtained from the 'self-consistency' condition of the applied decoration-iteration transformation (for more details see Ref.~\cite{Can06}). The mapping relationship~(\ref{eq:Z}) between the partition functions ${\cal Z}$ and ${\cal Z}_{\rm IC}$ completes the exact calculation of the partition function, since the partition function of the uniform spin-1/2 Ising chain can simply be calculated within the framework of the transfer-matrix method~\cite{Kra44,Bax82}. In the thermodynamic limit $N\to\infty$, one actually arrives at the following expression for the corresponding partition function
\begin{eqnarray}
\label{eq:ZIC}
{\cal Z}_{\rm IC}(\beta, R, H_{\rm IC}) &=& {\rm e}^{N\beta R/4}
\Big[\cosh(\beta H_{\rm IC}/2) \nonumber\\
&&{}+ \sqrt{\sinh(\beta H_{\rm IC}/2)+{\rm e}^{-\beta R}}\,\,
\Big]^{N}.
\end{eqnarray}
At this stage, exact results for all physical quantities, which are important for understanding of the magnetic behaviour of the investigated spin system, follow straightforwardly, whereas the Helmholtz free energy of the spin-1/2 Ising-Heisenberg diamond chain may be connected to the Helmholtz free energy of the uniform spin-1/2 Ising chain (${\cal F}_{\rm IC} = -k_{\rm B}T\ln{\cal Z}_{\rm IC})$ through the relation
\begin{eqnarray}
\label{eq:F}
{\cal F} = {\cal F}_{\rm IC} - N k_{\rm B}T\ln A.
\end{eqnarray}

\subsection{Magnetization and correlation functions}
\label{subsec:mag_cor}
The sublattice magnetization $m_{\rm I}$ and $m_{\rm H}$ reduced per one Ising and Heisenberg spin, respectively, can easily be calculated by differentiating Eq.~(\ref{eq:F}) with respect to the particular magnetic field acting on Ising and Heisenberg spins, respectively:
\begin{eqnarray}
m_{\rm I}\! &=& -\frac{1}{N}\!\left(\frac{\partial {\cal F}}{\partial H_{\rm I}}\right)_{\!\!T} \!\!\!= -\frac{1}{N}\left(\frac{\partial {\cal F}_{\rm IC}}{\partial H_{\rm IC}}\right)_{\!\!T} \!\!\! = m_{\rm IC},
\\
m_{\rm H}\! &=& -\frac{1}{2N}\!\left(\frac{\partial {\cal F}}{\partial H_{\rm H}}\right)_{\!\!T} \!\!\! = \frac{{\cal J}_{H_{\rm H}}^1}{2}\!\left(\!\frac{1}{4}\!+\!m_{\rm IC}\!+\!\varepsilon_{\rm IC}\!\right) \nonumber\\
&&{}+\frac{{\cal J}_{H_{\rm H}}^2}{2}\!\left(\!\frac{1}{4}\!-\!m_{\rm IC}\!+\!\varepsilon_{\rm IC}\!\right)\!\! +\! \frac{{\cal J}_{H_{\rm H}}^3}{2}\!\left(\!\frac{1}{2}\!-\!2\varepsilon_{\rm IC}\!\right)\!.
\label{eq:mH}
\end{eqnarray}
In the above, the parameters $m_{\rm IC}$ and $\varepsilon_{\rm IC}$ denote the single-site magnetization and the correlation
function between the nearest-neighbour spins of the uniform spin-1/2 Ising chain~\cite{Kra44,Bax82} and the coefficients ${\cal J}_{H_{\rm H}}^{i}$ ($i=1,2,3$) mark the expressions ${\cal J}_{H_{\rm H}}^{i}=\frac{\partial}{\partial_{H_{\rm H}}}\ln G_i$
with $G_1 = G(\beta J_{\rm I}, \beta K/4)$, $G_2 = G(-\beta J_{\rm I}, \beta K/4)$ and $G_3 = G(0, -\beta K/4)$. In view of this notation, the total magnetization normalized per one spin of the diamond chain can be expressed as
\begin{eqnarray}
\label{eq:m}
m = \frac{1}{3}(m_{\rm I} + 2m_{\rm H}).
\end{eqnarray}
It is worthwhile to mention that both sublattice magnetization listed above can also be calculated by combining the mapping relation~(\ref{eq:Z}) with the exact mapping theorems developed by Barry {\it et al.}~\cite{Bar88,Kha90,Bar91} and the generalized Callen-Suzuki spin identity~\cite{Cal63,Suz65,Bal02}. This alternative approach is more general, since it enables to evaluate the ensemble average for any
combination of random spin variables involved in the total Hamiltonian of the system. For example, one easily obtains the exact expressions for other important physical quantities such as the pair correlation functions $C_{\rm II}^{zz}\equiv\langle\hat{\sigma}_{3k-2}^z\rangle$, $C_{\rm HH}^{zz}\equiv\langle\hat{S}_{3k-1}^z\hat{S}_{3k}^z\rangle$, $C_{\rm HH}^{xx}\equiv\langle\hat{S}_{3k-1}^x\hat{S}_{3k}^x\rangle$, and $C_{\rm IH}^{zz}\equiv\langle\hat{\sigma}_{3k-2}^z\hat{S}_{3k-1}^z\rangle$ as well as the four-spin correlation function $Q_{\rm IH}^{zz}\equiv\langle\hat{\sigma}_{3k-2}^z\hat{\sigma}_{3k+1}^z\hat{S}_{3k-1}^z\hat{S}_{3k}^z\rangle$ in this way:
\begin{eqnarray}
C_{\rm II}^{zz} &=& \varepsilon_{\rm IC},
\label{eq:cIz}
\\
C_{\rm HH}^{zz} &=& {\cal K}_1^{+}/4 + {\cal K}_1\,m_{\rm IC} + {\cal K}_1^{-}\,\varepsilon_{\rm IC},
\label{eq:cHz}
\\
C_{\rm HH}^{xx} &=& {\cal K}_2^{+}/4 + {\cal K}_2\,m_{\rm IC} + {\cal K}_2^{-}\,\varepsilon_{\rm IC},
\label{eq:cHx}
\\
C_{\rm IH}^{zz} &=& {\cal K}_3/2 + ({\cal K}_3^{+}+{\cal K}_3^{-})\,m_{\rm IC} + 2{\cal K}_3\,\varepsilon_{\rm IC}
\label{eq:cIHz}
\\
Q_{\rm IH}^{zz} &=& {\cal K}_1^{-}/16 + {\cal K}_1m_{\rm IC}/4 + {\cal K}_1^{+}\varepsilon_{\rm IC}/4
\label{eq:qIHz}
\end{eqnarray}
The symbol $\langle\ldots\rangle$ denotes the standard canonical average performed over the ensemble defined on the investigated Ising-Heisenberg diamond chain and the coefficients ${\cal K}_i^{\pm}$ and ${\cal K}_i$ ($i=1,2,3$) emerging in Eqs.~(\ref{eq:cHz})--(\ref{eq:qIHz}) mark the expressions ${\cal K}_i^{\pm} = F_i(\beta J_{\rm I}, \beta K/4) + F_i(-\beta J_{\rm I}, \beta K/4) \pm 2F_i(0, -\beta K/4)$ and ${\cal K}_i = F_i(\beta J_{\rm I}, \beta K/4) - F_i(-\beta J_{\rm I}, \beta K/4)$ with
\begin{eqnarray}
F_1(x,y) &=& -[{\rm e}^{-\beta J_{\rm H}/4 - y/4}\cosh(x - \beta H_{\rm H}) \nonumber\\
&&-{\rm e}^{\beta J_{\rm H}/4 + y/4}\!\cosh(\beta J_{\rm H}\Delta/2)]\big/\,[2G(x,y)],\nonumber\\
F_2(x,y) &=& -{\rm e}^{\beta J_{\rm H}/4 + y/4}\sinh(\beta J_{\rm H}\Delta/2)\big/\,[2G(x,y)],\nonumber\\
F_3(x,y) &=& -{\rm e}^{-\beta J_{\rm H}/4 - y/4}\sinh(x - \beta H_{\rm H})\big/\,[4G(x,y)]\nonumber.
\end{eqnarray}

\subsection{Thermodynamics}
\label{subsec:thermo}

Before concluding this section, it is worthy of notice that several basic thermodynamic quantities such as the susceptibility $\chi$ and the specific heat ${\cal C}$ can also be readily derived from the Helmholtz free energy~(\ref{eq:F}) by using the standard thermodynamic relations
\begin{eqnarray}
\label{eq:SC}
\chi = -\!\left(\frac{\partial^2 {\cal F}}{\partial H^2}\right)_{\!\!T}\,\,,\qquad
{\cal C} = -T\!\left(\frac{\partial^2 {\cal F}}{\partial T^2}\right)_{\!\!H}\,\,.
\end{eqnarray}
However, the final expressions for the quantities $\chi$ and ${\cal C}$ are too cumbersome, therefore we do not write them here explicitly.

\section{Results and discussion}
\label{sec:results}
In this section, we will focus our attention to the most interesting results to be obtained for ground state and basic thermodynamic quantities of the investigated diamond chain. Even though all the results derived in the preceding section are general and hold regardless of whether ferromagnetic or antiferromagnetic exchange interactions are assumed, we will consider both the exchange parameters $J_{\rm I}$ and $J_{\rm H}$ to be antiferromagnetic ($J_{\rm I}>0, J_{\rm H}>0$) only, since it can be expected that the magnetic behaviour of the model with the antiferromagnetic interactions in the external longitudinal magnetic field should be much more interesting compared with its ferromagnetic counterpart. Moreover, our subsequent analysis will be restricted to the special case with the uniform magnetic field $H_{\rm H} = H_{\rm I} = H$ acting on the spin-$1/2$ atoms, which physically corresponds to the situation with the equal $g$-factors for both kinds of the magnetic atoms. Under these assumptions, we can set the Ising interaction $J_{\rm I}$ as the energy unit and introduce the following set of dimensionless parameters: $\alpha = J_{\rm H}/J_{\rm I}$, $\alpha_4 = K/J_{\rm I}$, $h = H/J_{\rm I}$, and $t = k_{\rm B}T/J_{\rm I}$, as describing the strength of the Heisenberg interaction normalized with respect to the Ising interaction, the strength of the four-spin interaction normalized with respect to the Ising interaction, the relative strength of the external magnetic field and the dimensionless temperature, respectively.

\subsection{Ground-state phase diagrams}
\label{subsec:GS}
\begin{figure}[htb]%
\vspace{-0.7cm}\hspace{-0.7cm}
\includegraphics*[width=1.16\linewidth,height=0.85\linewidth]{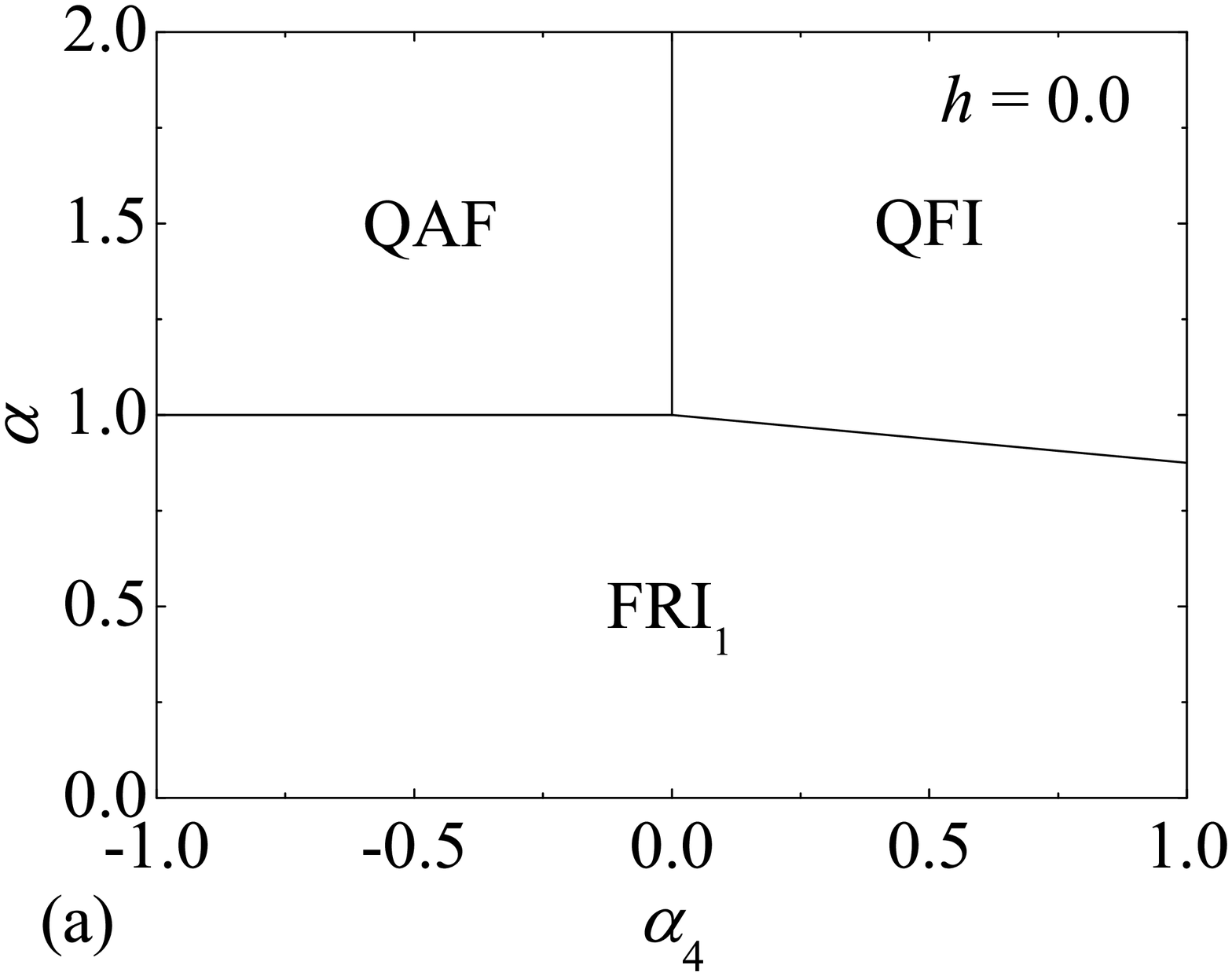}
\vspace{-1.5cm}
\\
\vspace{-1.0cm}\hspace{-0.7cm}
\includegraphics*[width=1.16\linewidth,height=0.85\linewidth]{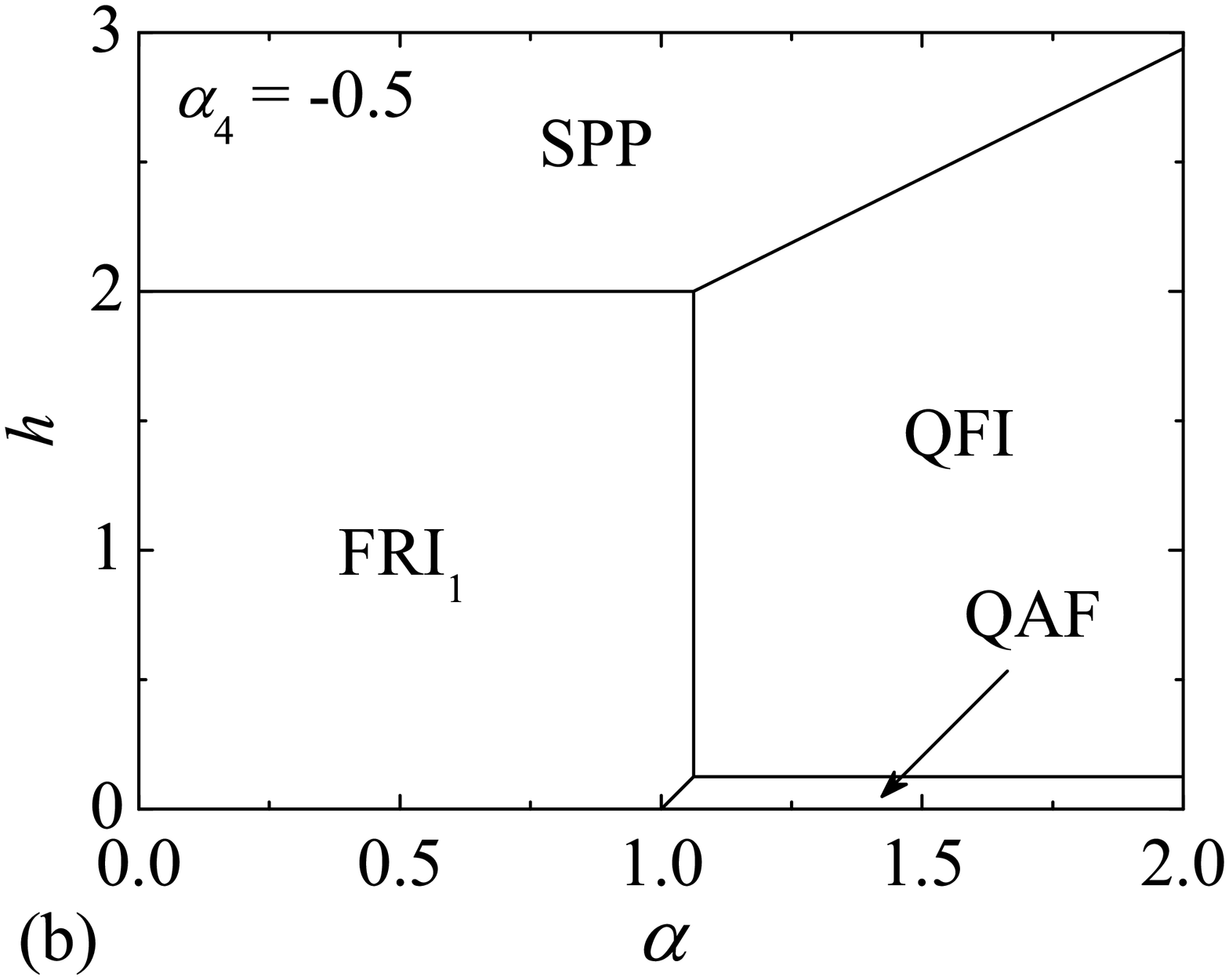}
\\
\vspace{-1.0cm}\hspace{-0.7cm}
\includegraphics*[width=1.16\linewidth,height=0.85\linewidth]{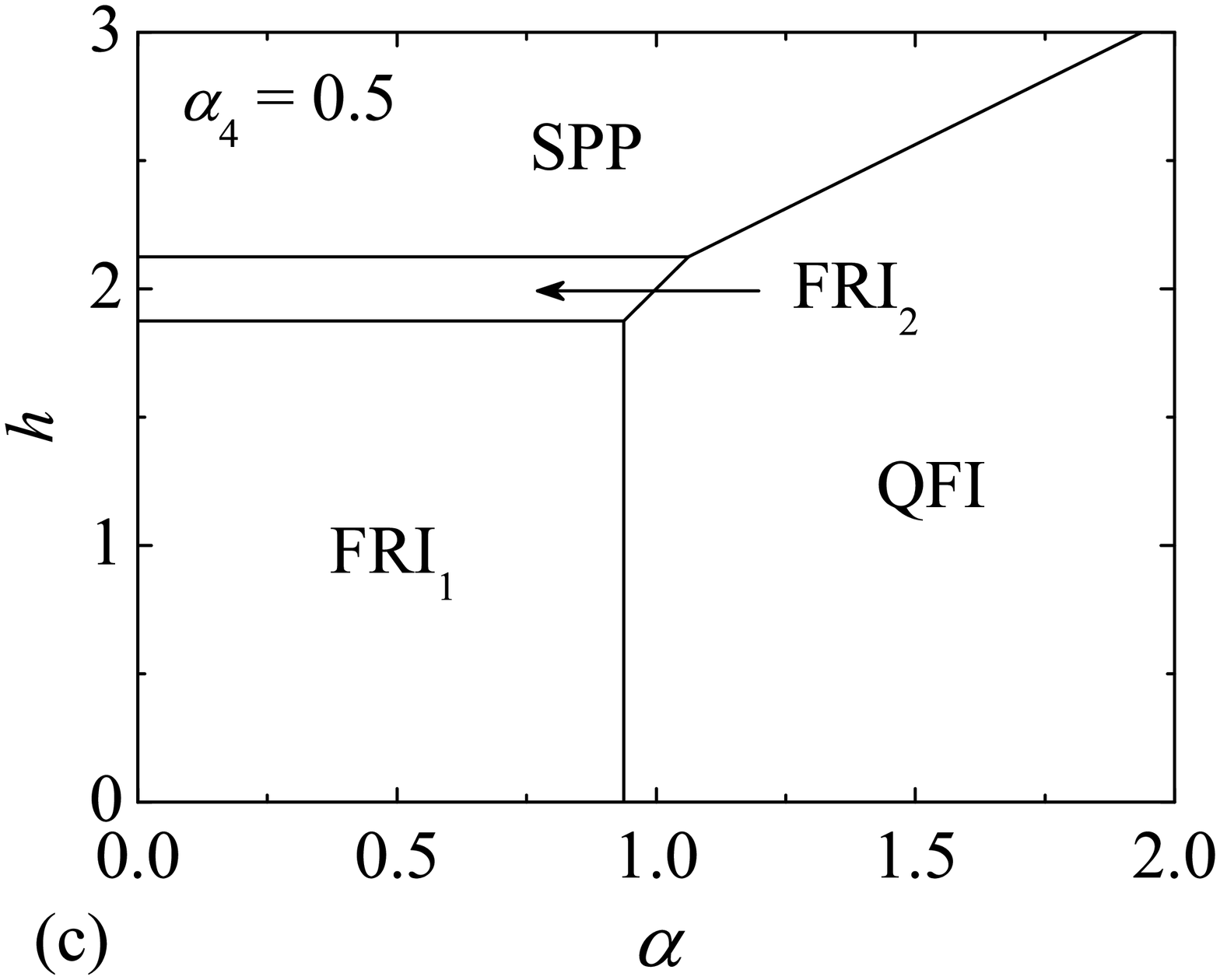}
\vspace{0.5cm}
\caption{%
Ground-state phase diagrams of the antiferromagnetic spin-$1/2$ Ising-Heisenberg diamond chain with the four-spin interaction (a) for the system without the external magnetic field and (b),(c) for the system in the external magnetic field with the fixed $\alpha_4=-0.5$, $0.5$, respectively, when $\Delta=1$.}
\label{fig2}
\end{figure}

First, let us construct the ground-state phase diagrams for the antiferromagnetic spin-$1/2$ Ising-Heisenberg diamond chain with the four-spin interaction under the assumption of the fixed exchange anisotropy parameter $\Delta=1$, which clarify the magnetic behaviour of the system without the external magnetic field and the magnetic behaviour of the system in the external magnetic field.
The ground state phase diagram for the spin-$1/2$ Ising-Heisenberg diamond chain with the four-spin interaction without the external magnetic field is illustrated in Fig.~\ref{fig2}(a). As one can see, three different phases FRI$_{1}$, QFI and QAF appear in the ground state of the system without the external magnetic field due to a mutual interplay between the exchange interactions $J_{\rm H}$, $J_{\rm I}$, $K$ and the fixed exchange anisotropy $\Delta=1$. They spin arrangements can be unambiguously characterized as follows:
\begin{eqnarray}
\label{eq:FRI_1}
&&|{\rm FRI_1}\rangle = \prod_{k=1}^{N}|-\rangle_{3k-2}|+, +\rangle_{3k-1,3k}, \nonumber \\
&&m_{\rm I} = -1/2,\, m_{\rm H} = 1/2, C_{\rm II}^{zz} = 1/4, \nonumber \\
&&C_{\rm HH}^{zz} = 1/4,\, C_{\rm HH}^{xx} = 0,\, C_{\rm IH}^{zz} = -1/4,\, Q_{\rm IH}^{zz} = 1/16, \nonumber \\
&&{\cal E}_{\rm FRI_1} = J_{\rm H}/4 - J_{\rm I} + K/16 - H/2\,;
\end{eqnarray}
\begin{eqnarray}
\label{eq:QFI}
&&|{\rm QFI}\rangle = \prod_{k=1}^{N}|+\rangle_{3k-2}
\frac{1}{\sqrt{2}}\Big(|+, -\rangle - |-, +\rangle\Big)_{3k-1,3k} , \nonumber \\
&&m_{\rm I} = 1/2,\, m_{\rm H} = 0, C_{\rm II}^{zz} = 1/4, \nonumber \\
&&C_{\rm HH}^{zz}= -1/4,\, C_{\rm HH}^{xx} = -1/4,\, C_{\rm IH}^{zz} = 0,\, Q_{\rm IH}^{zz} = -1/16,
\nonumber \\
&&{\cal E}_{\rm QFI} = -J_{\rm H}/4 -J_{\rm H}\Delta/2 - K/16 - H/2\,;
\end{eqnarray}
\begin{eqnarray}
\label{eq:QAF}
&&|{\rm QAF}\rangle = \prod_{k=1}^{N}|(-)^{3k-2}\rangle_{3k-2}
\frac{1}{\sqrt{2}}\Big(|+, -\rangle - |-, +\rangle\Big)_{3k-1,3k} , \nonumber \\
&&m_{\rm I} \,\,\,= 0,\, m_{\rm H} = 0, C_{\rm II}^{zz} = -1/4, \nonumber   \\
&&C_{\rm HH}^{zz} = -1/4,\, C_{\rm HH}^{xx} = -1/4,\, C_{\rm IH}^{zz} = 0,\, Q_{\rm IH}^{zz} = 1/16, \nonumber \\
&&{\cal E}_{\rm QAF} = - J_{\rm H}/4 - J_{\rm H}\Delta/2 + K/16\,.
\end{eqnarray}
In above, the ket vectors $|\pm\rangle$ after relevant products describe the spin states $\sigma^z = \pm 1/2$ and $S^z = \pm 1/2$ of the Ising and Heisenberg spins, respectively. The wave function of the phase QAF is written with the help of the expression $(-)^{3k-2}\in\{+,-\}$, where $k = 1, ..., N$. Obviously, all three phases meet at one triple point given by the condition
\begin{eqnarray}
\label{eq:T}
{\rm T}=[\alpha_4, \alpha] = \left[0, \frac{2}{1+\Delta}\right]\!.
\end{eqnarray}
As one clearly sees from Eqs.~(\ref{eq:FRI_1}),
the phase FRI$_1$ represents the semi-classically ordered ferrimagnetic phase with the antiparallel alignment between the nearest-neighbouring Ising and Heisenberg spins. By contrast, in other two phases QFI and QAF one observes the effect of quantum fluctuations on Heisenberg bonds. Namely, the pairs of decorating Heisenberg spins reside at a quantum superposition of spin states described by the antisymmetric wave function $\big(|+, -\rangle - |-, +\rangle\big)_{3k-1,3k}/\sqrt{2}$ in both the phases QFI and QAF. In the former phase QFI, all nodal Ising spins occupy the spin state $\sigma^z = 1/2$, while in the latter one QAF, one finds a perfect antiferromagnetic order in the Ising sublattice. Moreover, one may conclude from the location of QAF in phase diagrams shown in Figs.~\ref{fig2}(a) and (b) that this phase is a result of a mutual interplay between the quantum fluctuations arising from the antiferromagnetic Heisenberg interaction $J_{\rm H}$ and the ferromagnetic four-spin interaction $K$. Actually, the phase QAF appears in the ground state only when $\alpha>1$ and $\alpha_{4}<0$, regardless of whether the external magnetic field $h$ is zero or non-zero. Furthermore, the phase diagram plotted in Fig.~\ref{fig2}(c) suggests that in response to the applied magnetic field $h$ there also may arise the ferrimagnetic phase FRI$_2$ and the saturated paramagnetic phase SPP in the ground state besides the aforementioned FRI$_1$ , QFI and QAF phases:
\begin{eqnarray}
\label{eq:FRI_2}
&&|{\rm FRI_2}\rangle = \prod_{k=1}^{N}|(-)^{3k-2}\rangle_{3k-2}|+, +\rangle_{3k-1,3k}, \nonumber \\
&&m_{\rm I} \,\,\,= 0,\, m_{\rm H} = 1/2, C_{\rm II}^{zz} = -1/4,
\nonumber\\
&&C_{\rm HH}^{zz} = 1/4,\, C_{\rm HH}^{xx} = 0,\, C_{\rm IH}^{zz} = 0,\, Q_{\rm IH}^{zz} = -1/16,\\
&&{\cal E}_{\rm FRI_2} = J_{\rm H}/4 - K/16 - H\,;
\nonumber
\end{eqnarray}
\begin{eqnarray}
\label{eq:SPP}
&&|{\rm SPP}\rangle = \prod_{k=1}^{N}|+\rangle_{3k-2}
|+, +\rangle_{3k-1,3k}, \nonumber \\
&&m_{\rm I} \,\,\,= 1/2,\, m_{\rm H} = 1/2, C_{\rm II}^{zz} = 1/4, \nonumber \\
&&C_{\rm HH}^{zz} = 1/4,\, C_{\rm HH}^{xx} = 0,\, C_{\rm IH}^{zz} = 1/4,\, Q_{\rm IH}^{zz} = 1/16,
\\
&&{\cal E}_{\rm SPP} = J_{\rm H}/4 + J_{\rm I} + K/16 - 3H/2\,.
\nonumber
\end{eqnarray}
It can easily be understood from Eqs.~(\ref{eq:FRI_2}) that in the FRI$_2$ state, the pairs of Heisenberg spins occupy the ferromagnetic state $|+, +\rangle_{3k-1,3k}$, while the nodal Ising spins are ordered antiferromagnetically against each other. It is clear that this phase appears as a result of a mutual interplay between applied magnetic field $h$ and the antiferromagnetic four-spin interaction $K$, since it is stable in the ground state only when $h>1$ and $\alpha_{4}>0$ [see Fig.~\ref{fig2}(c)]. Finally, as one can expect, the system always undergoes a phase transition towards the fully saturated paramagnetic phase SPP, where all Ising and Heisenberg spins are completely aligned towards the external-field direction [see Eqs.~(\ref{eq:SPP})] in high magnetic fields.

\subsection{Magnetization process}
\label{subsec:MvsH}
\begin{figure*}[htb]%
\vspace{-0.45cm}\hspace{0.8cm}
\includegraphics*[width=0.925\linewidth,height=1.23\linewidth]{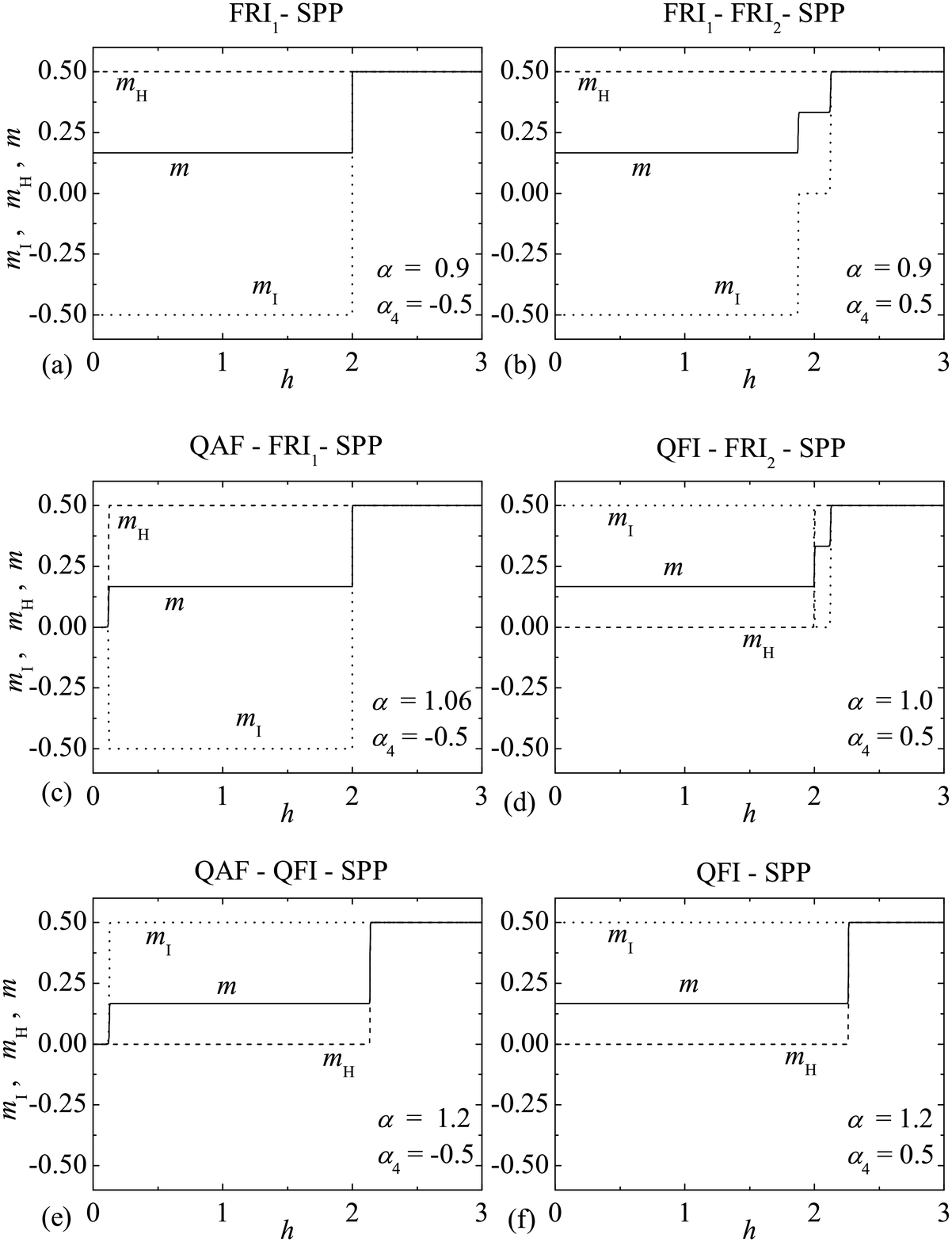}
\vspace{0.0cm}
\caption{%
The zero-temperature total and sublattice magnetization against the external magnetic field for the fixed four-spin interaction $\alpha_4=-0.5$ (left column), $\alpha_4=0.5$ (right column) and several values of the interaction parameter $\alpha$.}
\label{fig3}
\end{figure*}
Next, let us turn our attention to the magnetization process of the investigated diamond chain. For this purpose, the total and sublattice magnetization against the external magnetic field are plotted in Fig.~\ref{fig3} for several values of the interaction parameters $\alpha$ and $\alpha_4$. In agreement with the ground-state phase diagrams shown in Figs.~\ref{fig2}(b) and (c), the plotted zero-temperature magnetization curves reflect altogether six different sequences of field-induced phase transitions: FRI$_1$--SPP, QAF--FRI$_1$--SPP, QAF--QFI--SPP, FRI$_1$--FRI$_2$--SPP, QFI--FRI$_2$--SPP and QFI--SPP. Moreover, it is clear from Fig.~\ref{fig3} that one and/or two intermediate plateaus can be found in magnetization curves before the total magnetization $m$ tends towards its saturation value. It is worthwhile to remark that the identified magnetization plateaus, appearing at 0, 1/3 and 2/3 of the saturation magnetization when normalizing the total magnetization with respect to its saturation value, satisfy the Oshikawa-Yamanaka-Affeck rule $p(S_{u}-m)\in {\rm integer}$~\cite{Osh97}, which has been proposed as a necessary condition for the formation of quantized plateaus ($p$ is a period of the ground state, $S_u$ and $m$ are the total spin and total magnetization of the elementary unit).

\subsection{Thermodynamics}
\label{subsec:thermo}
\begin{figure}[htb]%
\vspace{-0.75cm}\hspace{-1.0cm}
\includegraphics*[width=1.25\linewidth,height=0.95\linewidth]{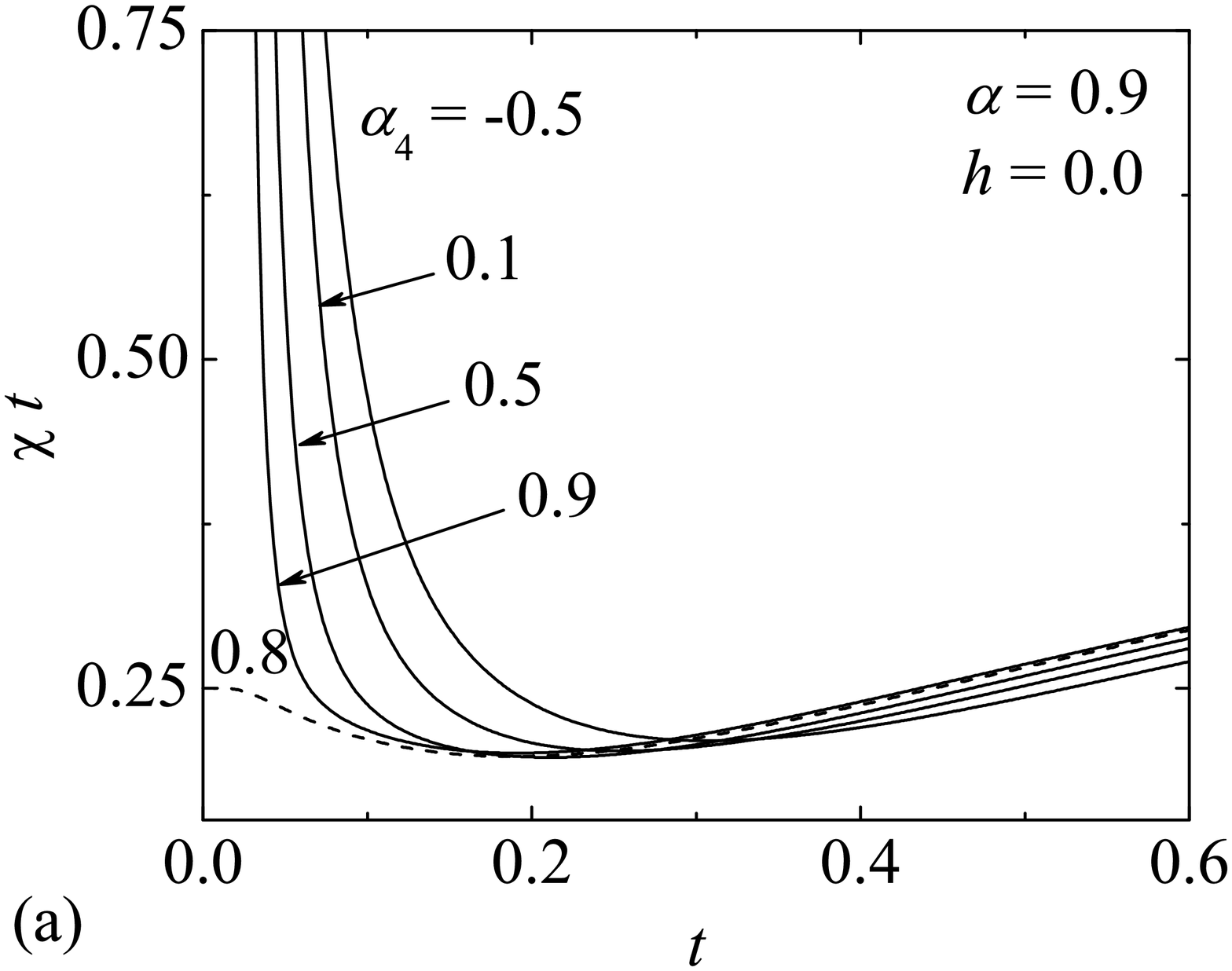}
\vspace{-1.0cm}
\\
\vspace{-0.5cm}\hspace{-1.0cm}
\includegraphics*[width=1.25\linewidth,height=0.95\linewidth]{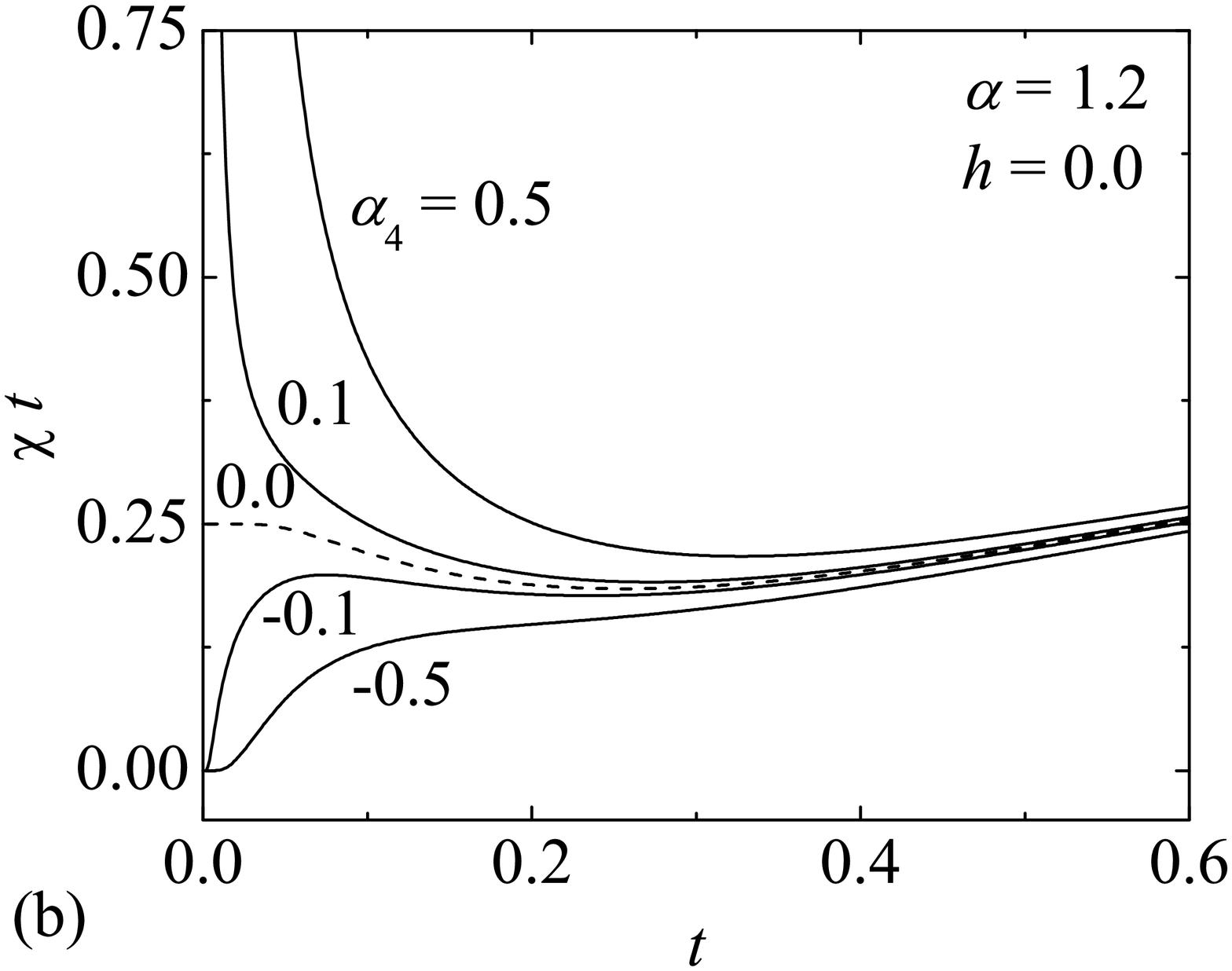}
\vspace{-0.25cm}
\caption{%
The magnetic susceptibility multiplied by the temperature as a function of the temperature in the zero magnetic field for several values of the four-spin interaction $\alpha_4$ when (a)~$\alpha=0.9$ and (b)~$\alpha=1.2$.}
\label{fig4}
\end{figure}

Next, let us turn our attention to temperature dependencies of some thermodynamic quantities. Figure~\ref{fig4} depicts the magnetic susceptibility multiplied by the temperature $(\chi t)$ as a function of the temperature in the zero field for several values of the four spin interaction $\alpha_4$ when $\alpha=0.9$ and $1.2$. As can be seen, the thermal dependencies of the product $\chi t$ either look like those that are typical for 1D quantum ferrimagnets~\cite{Yam99,Nak02,Yam04} or have an antiferromagnetic character. Actually, when the values of the four spin interaction $\alpha_4$ and the interaction parameter $\alpha$ are chosen so that the phases FRI$_1$ or QFI constitute the ground state, then the quantity exhibits a round minimum upon cooling and then exponentially diverges under further temperature suppression [see for instance the curves for $\alpha_4=-0.5, 0.1, 0.9$ in Fig.~\ref{fig4}(a) and for $\alpha_4=0.1, 0.5$ in Fig.~\ref{fig4}(b)]. These dependencies clearly reveal a ferromagnetic-antiferromagnetic crossover. The temperature variation of $\chi t$ is generally a monotonically decreasing function for ferromagnets and monotonically increasing function for antiferromagnets as the temperature increases. In this respect, the marked low-temperature divergence of $\chi t$ emerges because of the ferromagnetic excitations of spins from the magnetically ordered ground state, while the monotonous increase of $\chi t$ observed at higher temperatures is the result of the temperature-induced antiferromagnetic spin excitations. On the other hand, the quantity $\chi t$ always exponentially tends to zero in the limit $t\rightarrow0$ when the interaction parameters $\alpha_4$ and $\alpha$ are chosen so that the phase QAF constitutes the ground state [see the curves for $\alpha_4=-0.1$ and $-0.5$ in Fig.~\ref{fig4}(b)]. Finally, other notable thermal variations of the magnetic susceptibility multiplied by the temperature occur for the boundary values of the four-spin interaction $\alpha_4=0.8$ and $0.0$ that correspond to the phase boundaries FRI$_1$--QFI and QAF--QFI in the ground state, respectively. Here, the magnetic susceptibility diverges as $t^{-1}$ at low temperatures and the product $\chi t$ tends to the value $1/4$ in the zero-temperature limit, which can be explained in terms of the Curie law.

\begin{figure*}[htb]%
\vspace{-0.75cm}
\includegraphics*[width=1.0\textwidth]{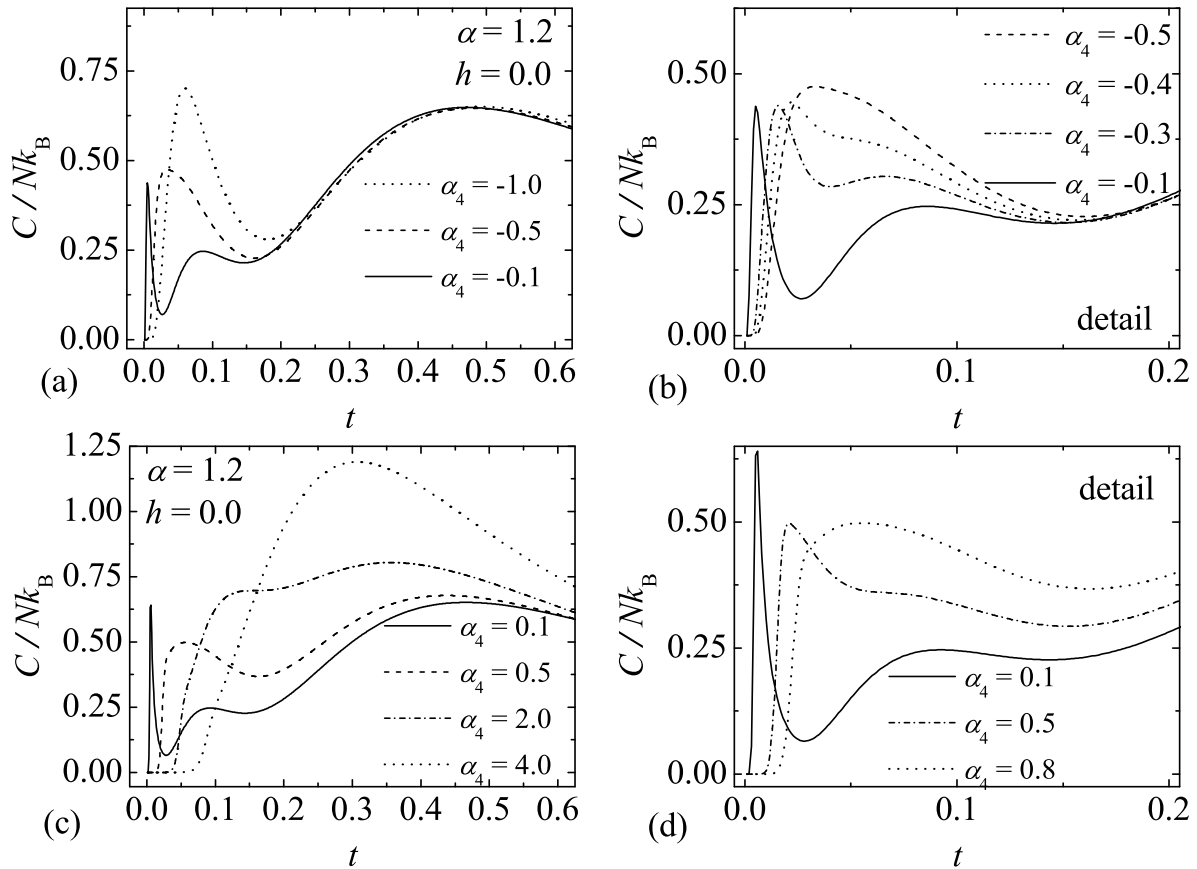}
\vspace{-0.75cm}
\caption{Temperature variations of the zero-field specific heat for $\alpha=1.2$ and several values of the four-spin interaction $\alpha_4$. The right column illustrates in detail the regions rather close to the triple point, where the phases FRI$_1$, QFI and QAF coexist in the ground state.}
\label{fig5}
\end{figure*}

Another quantity, which is important for understanding of thermodynamics, is the specific heat. Some temperature variations of the zero-field specific heat are plotted
in Fig.~\ref{fig5} for the fixed $\alpha=1.2$ and several values of the four-spin interaction $\alpha_4$. Note that values of the exchange parameters $\alpha$ and $\alpha_4$ are chosen so as to match the region near the triple point T given by Eq.~(\ref{eq:T}), where the phases FRI$_1$, QAF and QFI coexist in the ground state [see Fig.~\ref{fig2}(a)].
As one sees from Figs.~\ref{fig5}(a) and (c), if the four-spin interaction $\alpha_4$ is chosen to be sufficiently far from the triple point, then temperature dependencies of the specific heat with one or two round maxima are observable. Obviously, besides the usual Schottky-type maximum detected at high-temperature part of the specific heat, there also appears an additional second maximum located in the low-temperature part of these curves [see the curves labeled as $\alpha_4=-1.0$ in Fig.~\ref{fig5}(a) and as $\alpha_4=2.0$ in Fig.~\ref{fig5}(c)]. As expected, the low-temperature peak becomes more pronounced, the closer value of the four-spin interaction $\alpha_4$ is selected to the triple point [see e.g. the curves labeled as $\alpha_4=2.0$ and $0.5$ in Fig.~\ref{fig5}(c)].
These observations suggest that the double-peak structure in the specific heat curves originates from thermal excitations between the ground-state spin configuration and the ones close enough in energy to the ground state. Apart from these rather trivial findings, a remarkable triple-peak specific heat curve can also be found when the region rather close to the triple point is considered [see the curves labeled as $\alpha_4=-0.1$ in Fig.~\ref{fig5}(a) and $\alpha_4=0.1$ in Fig.~\ref{fig5}(c)]. Besides two afore-described peaks, there also occurs an additional third peak to be located at very low temperatures. There are strong indications that an appearance of this sharp maximum is the result of strong thermal excitations between the ground-state configuration of Ising spins in QAF and the ones in QFI and FRI$_1$. In accordance with this statement, the peculiar third maximum gradually shifts towards higher temperatures with retiring from the triple point until it coalesces with the second low-temperature peak [see Figs.~\ref{fig5}(b) and ~\ref{fig5}(d)]. Finally, it is worth noting that remarkable thermal dependencies of the specific heat with triple-peak structure can also be detected in the region sufficiently close to the triple point T when very small external fields are applied to the system. The similar triple-peak specific heat curves have been recently observed in the simpler version of this model without the four-spin interaction when applying small external field to the system driven by the spin frustration into the disordered ground state~\cite{Can06}.

\section{Concluding remarks}
\label{sec:conclusions}
In the present paper, the ground-state properties, magnetization process and thermodynamics of the symmetric antiferromagnetic spin-1/2 Ising-Heisenberg diamond chain with the Ising four-spin interaction has been particularly investigated within the framework of the generalized decoration-iteration mapping transformation. We have shown that the investigated diamond chain with the Ising four-spin interaction has five different ground states FRI$_1$, QFI, QAF, FRI$_2$ and SPP, providing the antiferromagnetic pair interactions (the Ising as well as the Heisenberg ones) and the fixed exchange anisotropy $\Delta = 1$. We have also found a rigorous evidence for one and/or two quantized plateaus in magnetization curves. Our further goal was to shed light on thermodynamics of the studied system. The results obtained for the magnetic susceptibility clearly demonstrate that the thermal dependencies of the zero-field susceptibility multiplied by the temperature either look like those that are typical for 1D quantum ferrimagnets or have an antiferromagnetic character. The most interesting result to emerge in the present study is the triple-peak structure in the specific heat curves that appears when assuming the ground-state region rather close to the triple point, where the phases FRI$_1$, QFI and QAF coexist.

Last but not the least, it should be mentioned that even though our theoretical investigation of the symmetric spin-1/2 Ising-Heisenberg diamond chain with the four-spin interaction has been mainly aimed at providing a deeper insight into the ground-state properties, magnetization process and thermal dependencies of the susceptibility and specific heat, it could be quite interesting to explore also an enhanced magnetocaloric effect in this diamond chain. Beside this, other particular case of this model with the ferromagnetic Heisenberg interaction ($J_{\rm H}<0$) deserves the attention, since there are some fundamental differences between magnetic behaviour of the hybrid Ising-Heisenberg models with distinct nature of the Heisenberg interaction (see Refs.~{\color{blue}\cite{Str02b,Can08,Jas08,Can10,Gal11}). Finally}, several further interesting extensions of the present version of the spin-1/2 Ising-Heisenberg diamond chain come also into question. For instance, it is possible to extend the present model by including the cyclic Heisenberg four-spin interaction between the Ising and Heisenberg spins of the diamond-shaped unit. In this direction will continue our next work.

\end{document}